\documentclass[fleqn,twoside]{article}
\usepackage{espcrc2}

\usepackage{graphicx}
\usepackage[figuresright]{rotating}

\newcommand{\AmS}{{\protect\the\textfont2
  A\kern-.1667em\lower.5ex\hbox{M}\kern-.125emS}}

\title{Topological susceptibility for the SU(3) Yang--Mills theory}

\author{Luigi Del Debbio\address[ldd]{CERN, Department of Physics, TH Division, 
                                       CH-1211 Geneva 23, Switzerland},
         Leonardo Giusti\address[lg]{Centre de Physique Th\'eorique, Case 907, CNRS
                                       Luminy, F-13288 Marseille Cedex 9, France}%
         \thanks{Speaker at the conference.}
         and Claudio Pica\address[cp]{Dipartimento di Fisica dell'Universit\`a di Pisa 
         and INFN, Via Buonarroti 2, I-56127 Pisa, Italy}}
       
\begin{document}

\begin{abstract}
We present the results of a computation of the topological susceptibility in 
the SU(3) Yang--Mills theory performed by employing the expression of the topological 
charge density operator suggested by Neuberger's fermions. In the continuum limit we 
find $r_0^4\chi = 0.059(3)$, which corresponds to $\chi=(191 \pm 5 \,\mathrm{MeV})^4$ 
if $F_K$ is used to set the scale. Our result supports the Witten--Veneziano 
explanation for the large mass of the $\eta'$.   
\vskip -0.25cm
\end{abstract}

\maketitle

\section{Introduction}\label{sec:intro}
The topological susceptibility in the Yang--Mills (YM) gauge
theory can be formally defined in Euclidean space-time as
\begin{equation}
\chi = \int d^4x\: 
\langle q(x) q(0) \rangle \; ,
\label{eq:chidef}
\end{equation}
where the topological charge density $q(x)$ is
\begin{equation}
q(x)=-\frac{1}{32\pi^2} 
\epsilon_{\mu\nu\rho\sigma} \mathrm{Tr}\Big[F_{\mu\nu}(x) F_{\rho\sigma}(x)\Big]\; .
\end{equation}
Besides its interest within the pure gauge theory, $\chi$
plays a crucial r\^ole in the QCD-based explanation of the large 
mass of the $\eta'$ meson proposed  by Witten and Veneziano (WV) 
\cite{Witten:1979vv,Veneziano:1979ec}. The WV mechanism
predicts that, at leading order in $N_\mathrm{f}/N_\mathrm{c}$,
the contribution due to the anomaly to the mass of the 
$U_{\mathrm{A}}(1)$ particle is 
given by \cite{Witten:1979vv,Veneziano:1979ec,Seiler:1987ig,Giusti:2001xh} 
\begin{equation}
\displaystyle \frac{F_{\pi}^2 m^2_{\eta^\prime}}{2 N_\mathrm{f}} = \chi\, ,  
\label{eq:WV}
\end{equation}
where $F_{\pi}$ is the pion decay constant. The discovery of a fermion 
operator \cite{Neuberger:1997bg} 
that satisfies the Ginsparg--Wilson (GW) relation \cite{Ginsparg:1982bj}
triggered a breakthrough in the 
understanding of the topological properties of the YM 
vacuum \cite{Hasenfratz:1998ri,Luscher:1998pq,Giusti:2001xh,Giusti:2004qd,Luscher:2004fu},
and made it possible to give a precise and unambiguous implementation
of the WV formula \cite{Giusti:2001xh}.
Indeed the naive lattice definition of the 
topological susceptibility has a finite continuum limit, which is 
universal \cite{Luscher:2004fu}, 
if the topological charge density, defined as suggested by GW fermions, is 
employed \cite{Giusti:2001xh,Giusti:2004qd,Luscher:2004fu}. 
This yields  the suggestive formula
\begin{equation}\label{eq:chil}
\chi = \lim_{\begin{array}{c}a\rightarrow 0\\V\rightarrow \infty\end{array}}
       \frac{\langle Q^2 \rangle}{V}
\end{equation}
with $Q \equiv \sum_x q(x) = n_+ - n_-$ being the topological charge,
$V$ the volume, and $n_+$ ($n_-$) the number of zero modes 
of $D$ with positive (negative) chirality in a given background. 
Using new simulation algorithms~\cite{Giusti:2002sm}, it is now possible to
investigate the WV scenario from first principles for the first time. 
More precisely, the aim of the work presented here, and 
fully described in \cite{DelDebbio:2004ns}, is to achieve an accurate and reliable determination 
of $\chi$ in the continuum limit, which in turn allows a verification
of the WV mechanism for 
the $\eta'$ mass. Several exploratory computations have 
already studied the susceptibility employing the GW definition of the topological 
charge~\cite{Edwards:1998wx,DeGrand:2002gm,Gattringer:2002mr,Hasenfratz:2002rp,Chiu:2003iw,Giusti:2003gf,DelDebbio:2003rn}. 
\begin{figure}
\begin{center}
\includegraphics*[width=7.0cm]{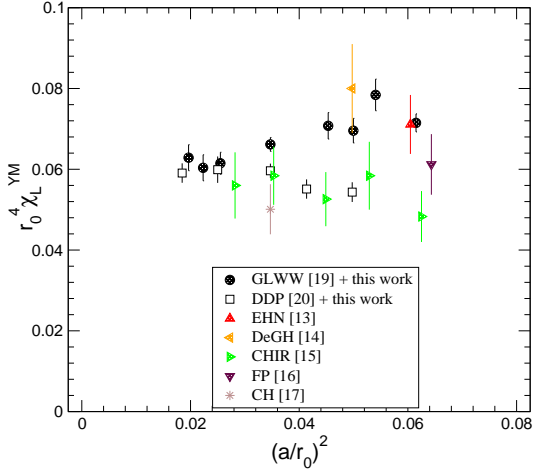}
\vskip -0.75cm
\caption{\label{fig:comparison} Comparison with other computations.}
\end{center}
\vskip -1.0cm
\end{figure}

\section{Lattice computation \label{sec:latt}}
The ensembles of gauge configurations are generated 
with the standard Wilson action and periodic boundary conditions,
using a combination of heat-bath and over-relaxation updates.
More details on the generation of the gauge configurations can be 
found in Refs.~\cite{Giusti:2003gf,DelDebbio:2003rn}. 
Table~\ref{tab:lattices} shows the list of simulated lattices, where
the bare coupling constant $\beta=6/g_0^2$, the linear size $L/a$ in
each direction and the number of independent configurations are
reported for each lattice. The topological charge density is 
defined as 
\begin{equation}\label{eq:qx}
q(x) = -\frac{\bar a}{2}\, \mathrm{Tr}\Big[\gamma_5 D(x,x)\Big] , 
\end{equation}
with $D$ being the massless Neuberger--Dirac operator 
\cite{Neuberger:1997bg}, and $s$ an adjustable parameter in the range $|s|<1$.
For a given gauge configuration, 
the topological charge is computed by counting the number of 
zero modes of $D$
(the interested reader should refer to \cite{DelDebbio:2004ns} for
more details). 
As $s$ is varied, $D$ defines a one-parameter family of fermion discretizations, 
which correspond to the same continuum theory but with different
discretization errors at finite lattice spacing. Our computation includes 
data sets computed for $s=0.4$ and $s=0.0$. A comparison with results 
previously obtained with various implementation of the Neuberger's operator is shown 
in Fig.~\ref{fig:comparison}.
\begin{table}[t]
\begin{center}
\setlength{\tabcolsep}{.25pc}
\begin{tabular}{llcrcc}
\hline
lat    &$\beta$&$L/a$&$N_{\mathrm{conf}}$& $\langle Q^2\rangle$ & $r_0^4\chi$ \\[0.125cm]
\hline
${\rm A}_1$&$6.0$   &$12$&$2452$&$1.633(48)$&$0.0654(22)$\\[0.125cm]
${\rm A}_2$&$6.1791$&$16$&$1138$&$1.589(76)$&$0.0629(32)$\\[0.125cm]
${\rm A}_3$&$5.8989$&$10$&$1460$&$1.737(72)$&$0.0696(30)$\\[0.125cm]
${\rm A}_4$&$6.0938$&$14$&$1405$&$1.535(63)$&$0.0615(27)$\\[0.125cm]
${\rm B}_0$&$5.8458$&$12$&$2918$&$5.61(16)$ &$0.0715(22)$\\[0.125cm]
${\rm B}_1$&$6.0$   &$16$&$1001$ &$5.58(28)$ &$0.0707(37)$\\[0.125cm]
${\rm B}_2$&$6.1366$&$20$&$963$  &$4.81(24)$ &$0.0604(32)$\\[0.125cm]
${\rm B}_3$&$5.9249$&$14$&$1284$ &$5.59(24)$ &$0.0708(33)$\\[0.125cm]
${\rm C}_0$&$5.8784$&$16$&$1109$ &$15.02(72)$&$0.0784(39)$\\[0.125cm]
${\rm C}_1$&$6.0$   &$20$&$931$  &$12.76(95)$&$0.0662(50)$\\[0.125cm]
${\rm D}$  &$6.0$   &$14$&$1577$ &$3.01(12)$ &$0.0651(27)$\\[0.125cm]
\hline
${\rm E}$&$5.9$   &$12$&$1349$&$2.79(12)$ &$0.0543(24)$\\[0.125cm]
${\rm F}$&$5.95$  &$12$&$1291$&$1.955(79)$&$0.0551(24)$\\[0.125cm]
${\rm G}$&$6.0$   &$12$&$3586$&$1.489(37)$&$0.0596(18)$\\[0.125cm]
${\rm H}$&$6.1$   &$16$&$962$ &$2.45(13)$ &$0.0599(33)$\\[0.125cm]
${\rm J}$&$6.2$   &$18$&$1721$&$2.114(76)$&$0.0591(24)$\\[0.125cm]
\hline
\end{tabular}
\caption{
\label{tab:lattices} Simulation parameters and results. For lattices ${\rm A}_1$--${\rm D}$
and ${\rm E}$--${\rm J}$, $s=0.4$ and $s=0.0$ respectively.}
\end{center}
\vskip -1.0cm
\end{table}
The
systematic uncertainties stem from finite-volume effects and from the
extrapolation needed to reach the continuum limit.
The pure gauge
theory has a mass gap, and therefore the topological susceptibility
approaches the infinite-volume limit exponentially fast with
$L$. Since the mass of the lightest glueball is around 1.5 GeV, finite-volume 
effects are expected to be far below our statistical errors as
soon as $L\ge 1$~fm. In order to further verify that no sizeable finite-volume
effects are present in our data, we simulated four lattices at
$\beta=6.0$ but with different linear sizes. The results obtained for $\chi$, reported in
Table~\ref{tab:lattices}, show no dependence on $L$ within our statistical errors.
\begin{figure}
\includegraphics*[width=7.0cm]{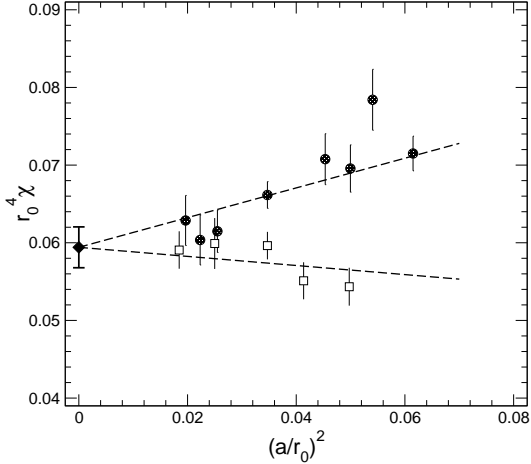}
\vskip -0.75cm
\caption{\label{fig:continuum} Continuum extrapolation of the adimensional
  product $r_0^4 \chi$. The $s=0.0$ and $s=0.4$ data sets are
  represented by black circles and white squares respectively. 
  The dashed lines represent the results of the
  combined fit described in the text. The filled diamond at $a=0$ is the
  extrapolated value in the continuum limit.}
\vskip -0.75cm
\end{figure}
At finite lattice spacing, $\chi$ is affected by discretization effects
starting at $O(a^2)$, which are not universal, and, in our case, depend
on the value of $s$ chosen to define the Neuberger operator. 
The values of the adimensional quantity $r_0^4 \chi$ that we obtain,
where $r_0/a$ is taken from~\cite{Guagnelli:1998ud}, 
are reported in Table~\ref{tab:lattices}. Data, displayed in Fig.~\ref{fig:continuum} as a 
function of $a^2/r_0^2$, show sizeable 
$O(a^2)$ effects for both the $s=0.4$ and $s=0.0$ samples. 
For $\beta \leq 6.0$, the difference
between the two discretizations is statistically significant.
A detailed data analysis indicate 
a linear dependence in $a^2$ of our results within  
our statistical errors \cite{DelDebbio:2004ns}.
\vspace{-0.25cm}
\section{Conclusions}\label{sec:phys}
A robust 
estimate of $\chi$ in the continuum limit can 
be obtained by performing a combined linear fit of the data. This fit gives
a very good value of $\chi^2_{\mathrm{dof}}$ when all sets are included, 
and is very stable if some points at larger values of $a^2/r_0^2$ are removed. 
In particular a combined fit of all points with $a^2/r_0^2<0.05$ gives 
$c_0=0.059(3)$ with $\chi^2_{\mathrm{dof}}\approx 0.73$.
Since $r_0$ is not directly accessible to experiments, we 
express our result in physical units by using the lattice determination
of $r_0 F_K=0.4146(94)$ in the quenched theory~\cite{Garden:1999fg} 
and we obtain
\begin{equation}
\chi = (191 \pm 5\, \mathrm{MeV})^4 \, ,
\end{equation}
which has to be compared with \cite{Veneziano:1979ec}
\begin{equation}\label{eq:exp}
\frac{F_\pi^2}{6} \Big(m^2_\eta + m^2_{\eta '} - 2 m^2_K\Big)\Big|_{\mathrm{exp}} \simeq 
(180\, \mathrm{MeV})^4 \; .
\end{equation}
Notice that Eq.~(\ref{eq:WV}) being only valid at the leading order in
a $N_{\mathrm{f}}/N_{\mathrm{c}}$ expansion, the ambiguity in the
conversion to physical units in the pure gauge theory is of the same
order as the neglected terms.  Our result supports the
Witten--Veneziano mechanism for explaining the large $\eta'$ mass.  A
measurement of the connected expectation values $\langle Q^{2n}
\rangle_c$ with the methods presented here would provide interesting
informations about the dependence of the free energy density on the
$\theta$ angle, or equivalently on the probability distribution of the
topological charge $P_Q$, putting on solid ground the results presented in
Ref.~\cite{DelDebbio:2002xa}. Unfortunately much higher statistics are
required in order to highlight the deviations from a Gaussian
distribution; higher momenta of the topological charge distribution
measured on our data are all compatible with zero within large
statistical errors. 


\begin{thebibliography}{10}
\expandafter\ifx\csname url\endcsname\relax
  \def\url#1{\texttt{#1}}\fi
\expandafter\ifx\csname urlprefix\endcsname\relax\def\urlprefix{URL }\fi
\providecommand{\eprint}[2][]{\url{#2}}

\bibitem{Witten:1979vv}
E.~Witten, Nucl. Phys. B156 (1979) 269.

\bibitem{Veneziano:1979ec}
G.~Veneziano, Nucl. Phys. B159 (1979) 213.

\bibitem{Seiler:1987ig}
E.~Seiler, I.O. Stamatescu, MPI-PAE/PTh 10/87.

\bibitem{Giusti:2001xh}
L.~Giusti et~al., Nucl. Phys. B628 (2002) 234.

\bibitem{Neuberger:1997bg}
H.~Neuberger, Phys. Rev. D57 (1998) 5417.

\bibitem{Ginsparg:1982bj}
P.H. Ginsparg, K.G. Wilson, Phys. Rev. D25 (1982) 2649.

\bibitem{Hasenfratz:1998ri}
P.~Hasenfratz, V.~Laliena, F.~Niedermayer, Phys. Lett. B427 (1998) 125.

\bibitem{Luscher:1998pq}
M.~Luscher, Phys. Lett. B428 (1998) 342.

\bibitem{Giusti:2004qd}
L.~Giusti et~al., Phys. Lett. B587 (2004) 157.

\bibitem{Luscher:2004fu}
M.~L\"uscher, hep-th/0404034.

\bibitem{Giusti:2002sm}
L.~Giusti et~al., Comput. Phys. Commun. 153 (2003) 31.

\bibitem{DelDebbio:2004ns}
L.~Del~Debbio, L.~Giusti, C.~Pica, hep-th/0407052.

\bibitem{Edwards:1998wx}
R.G. Edwards, U.M. Heller, R.~Narayanan, Phys. Rev. D59 (1999) 094510.

\bibitem{DeGrand:2002gm}
T.~DeGrand, U.M. Heller (MILC), Phys. Rev. D65 (2002) 114501.

\bibitem{Gattringer:2002mr}
C.~Gattringer, R.~Hoffmann, S.~Schaefer, Phys. Lett. B535 (2002) 358.

\bibitem{Hasenfratz:2002rp}
P.~Hasenfratz et~al., Nucl. Phys. B643 (2002) 280.

\bibitem{Chiu:2003iw}
T.W. Chiu, T.H. Hsieh, Nucl. Phys. B673 (2003) 217.

\bibitem{Giusti:2003gf}
L.~Giusti et~al., JHEP 11 (2003) 023.

\bibitem{DelDebbio:2003rn}
L.~Del~Debbio, C.~Pica, JHEP 02 (2004) 003.


\bibitem{Guagnelli:1998ud}
M.~Guagnelli, R.~Sommer, H.~Wittig (ALPHA), Nucl. Phys. B535 (1998) 389.

\bibitem{Garden:1999fg}
J.~Garden et~al. (ALPHA), Nucl. Phys. B571 (2000) 237.

\bibitem{DelDebbio:2002xa}
L.~Del Debbio, H.~Panagopoulos and E.~Vicari,
JHEP {\bf 0208} (2002) 044.
\end{thebibliography}

\end{document}